\documentclass{PoS}
\usepackage{natbib}
\setcitestyle{numbers}

\newcommand{\mnras}{MNRAS}
\newcommand{\apj}{ApJ}
\newcommand{\aj}{AJ}

\newcommand{\aap}{A\&A}
\newcommand{\araa}{ARA\&A}

\newcommand{\nat}{Nature}

\newcommand{\ssr}{SSRs}

\title{The MeerKAT International GHz Tiered Extragalactic Exploration (MIGHTEE) Survey}

\ShortTitle{The MIGHTEE Survey}

\author{\speaker{Matt\ J.\ Jarvis}$^{1,2}$, A.\ R.\ Taylor$^{2,3,4}$,
  I.\ Agudo$^{5}$, James\ R.\ Allison$^{6}$, R.\ P.\ Deane$^{7}$, B.\
  Frank$^{3,4}$, N.\ Gupta$^{8}$, I.\
  Heywood$^{1,7}$, N.\ Maddox$^{9}$, K.\ McAlpine$^{2,10}$, Mario\
 G.\ Santos$^{2}$, A.\ M.\ M.\  Scaife$^{11}$, M.\ Vaccari$^{2,12}$, J.\ T.\ L.\
  Zwart$^{3}$, E.\ Adams$^{9}$, D.\ J.\ Bacon$^{13}$, A.\ J.\
  Baker$^{14}$, Bruce.\ A.\ Bassett$^{15,16,17}$, P.\ N.\ Best$^{18}$, R.\
  Beswick$^{11}$, S.\ Blyth$^{3}$, Michael\ L.\ Brown$^{11}$, M.\
  Br\"{u}ggen$^{19}$, M.\ Cluver$^{2}$, S.\
  Colafranceso$^{20}$, G.\ Cotter$^{1}$, C.\ Cress$^{2,21}$, R.\ Dav\'e$^{2,16}$, C.\
  Ferrari$^{22}$, M.\ J.\
  Hardcastle$^{23}$, C.\ Hale$^{1}$, I.\ Harrison$^{11}$, P.\ W.\ Hatfield$^{1}$, H.-R.\
  Kl\"{o}ckner$^{24}$, S.\
  Kolwa$^{2,25}$, E.\ Malefahlo$^{2}$, T.\ Marubini$^{2}$, T.\ Mauch$^{10}$, K.\ Moodley$^{26}$, R.\
  Morganti$^{9}$, R.\ Norris$^{6,27}$, J.\ A.\ Peters$^{1}$, I.\ Prandoni$^{12}$, M.\ Prescott$^{2}$, S.\
  Oliver$^{28}$, N.\ Oozeer$^{10,15,29}$, H.\ J.\ A.\ R\"{o}ttgering$^{30}$, N.\
  Seymour$^{31}$, C.\
  Simpson$^{32}$, O.\ Smirnov$^{7}$, D.\ J.\ B.\ Smith$^{21}$, K.\
  Spekkens$^{33}$, J.\ Stil$^{34}$, C.\ Tasse$^{7,35}$, K.\ van der
  Heyden$^{3}$, I.\ H.\ Whittam$^{2}$, W.\ L.\ WIlliams$^{21}$\\
        $^{1}$University of Oxford \& $^{2}$University of the Western Cape\\
        E-mail: \email{matt.jarvis@physics.ox.ac.uk}}
\abstract{The MIGHTEE large survey project will survey four of the
  most well-studied extragalactic deep fields, totalling 20
  square degrees to $\mu$Jy sensitivity at Giga-Hertz frequencies, as well as an ultra-deep
  image of a single $\sim$1 deg$^2$ MeerKAT pointing. The observations will
provide radio continuum, spectral line and
  polarisation information. As such, MIGHTEE, along with the excellent
  multi-wavelength data already available in these deep fields, will allow a range of 
  science to be achieved. Specifically, MIGHTEE is designed to
  significantly enhance our understanding of, (i) the evolution of AGN
  and star-formation activity over cosmic time, as a function of
  stellar mass and environment, free of dust obscuration; (ii) the
  evolution of neutral hydrogen in the Universe and how this neutral
  gas eventually turns into stars after moving through the
  molecular phase, and how efficiently this can fuel
  AGN activity; (iii) the properties of cosmic magnetic fields and how they
  evolve in clusters, filaments and galaxies. 
  MIGHTEE will reach similar depth to the planned SKA all-sky survey, and
  thus will provide a pilot to the cosmology experiments that
  will be carried out by the SKA over a much larger survey volume. 
}

\FullConference{MeerKAT Science: On the Pathway to the SKA,\\
	25-27 May, 2016,\\
		Stellenbosch, South Africa}

\begin{document}

\section{Introduction}
The MeerKAT International GHz Tiered Extragalactic Exploration
(MIGHTEE) survey is a project being undertaken by an international
collaboration of researchers to explore cosmic evolution by 
creating deep images of the GHz radio emission in
continuum, spectral line and polarisation. The survey will be
conducted over 20 deg$^2$ of the best studied regions of the
extragalactic sky observable from the southern hemisphere, namely
COSMOS, XMM-LSS, ECDFS and ELAIS-S1. The nominal sensitivity will be
$\sim$1\,$\mu$Jy over the full bandwidth of 900-1670\,MHz, at a
resolution of $\sim$6\,arcsec, with additional observations made over
a smaller area with the S-band receiver.
In the following we outline the key science aims of MIGHTEE.

\section{Science Goals}

\subsection{MIGHTEE--LADUMA synergy: The cosmic evolution of H{\sc i}}
\begin{figure*}
\centering
\includegraphics[width=0.45\textwidth]{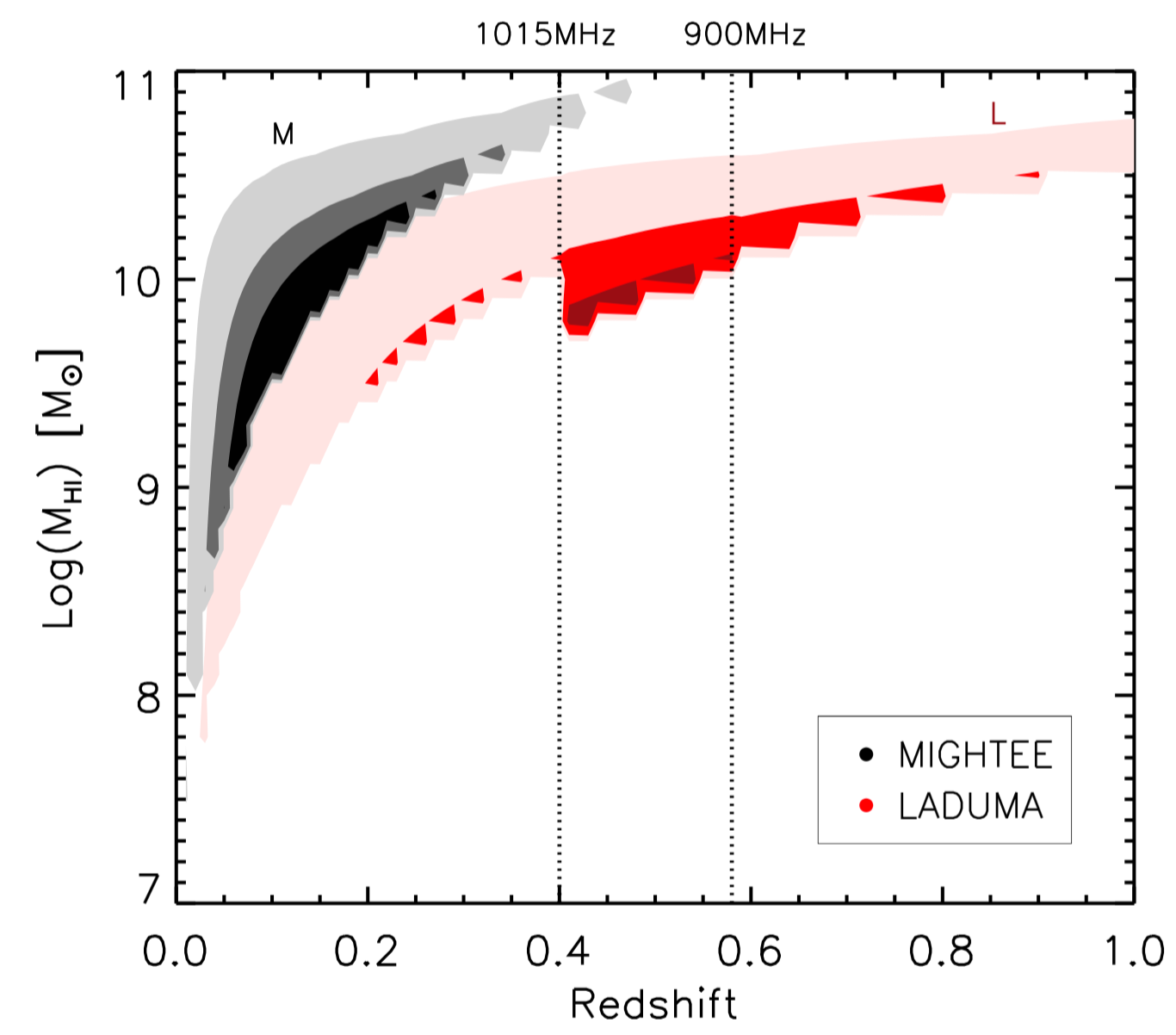}
\includegraphics[width=0.45\textwidth]{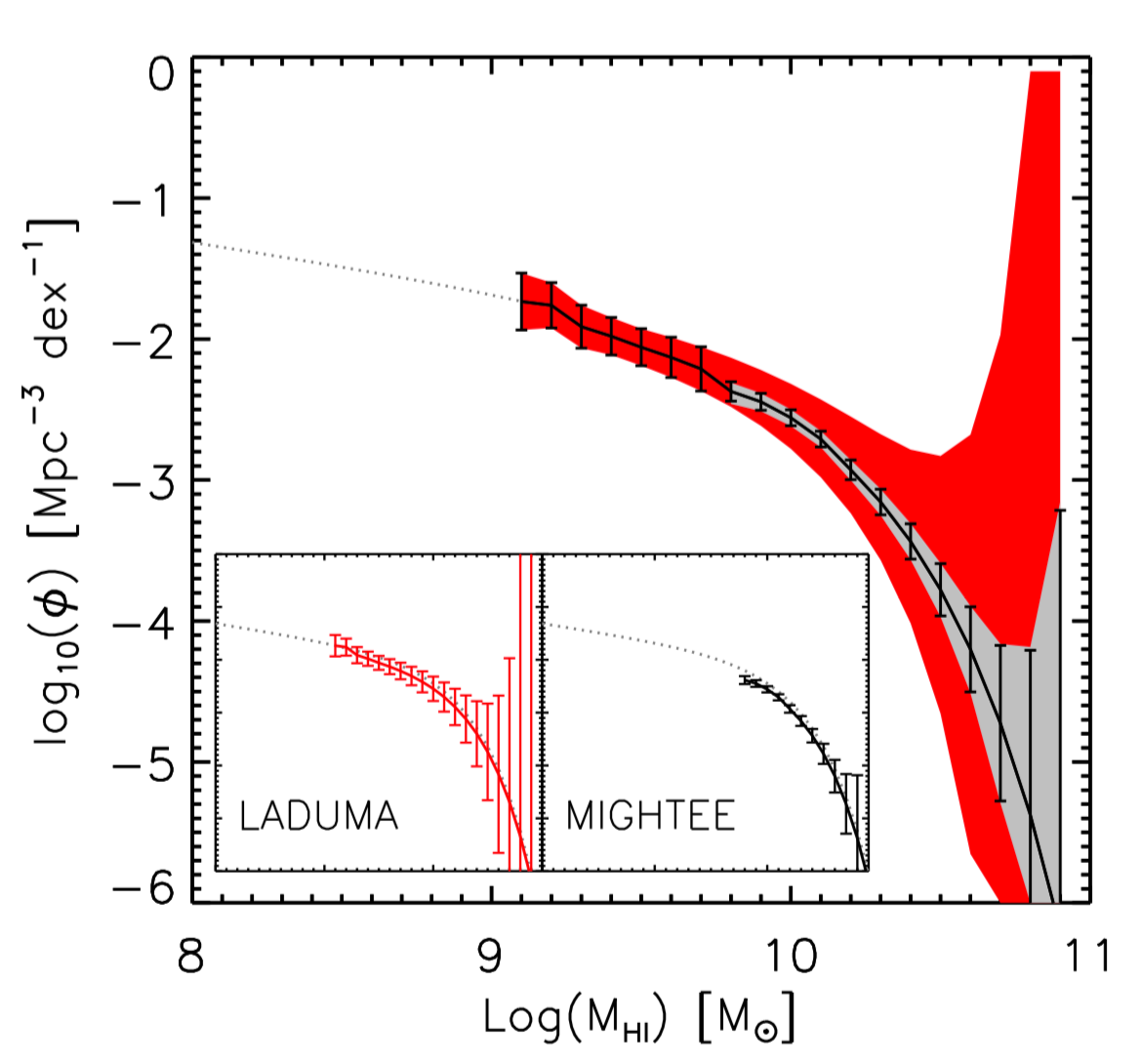}
\caption{\label{fig:HIMF}({\em left}) The expected coverage of the
  H{\sc i} mass versus redshift plane by combining the LADUMA and MIGHTEE surveys. The grey shaded regions show the MIGHTEE coverage and the red shaded regions show the LADUMA coverage. ({\em right}) The expected constraints on the H{\sc i} mass function. It is clear that MIGHTEE is required to measure the high-mass end whereas LADUMA pins down the low-mass end at $z<0.5$. Taken from \cite{Maddox2016}.}
\end{figure*}

The MIGHTEE H{\sc i} survey component and the LADUMA H{\sc i} survey
(Baker et al. these proceedings) can be thought of as two tiers of a
survey `wedding cake' at L-band. The $\sim$20 square degree MIGHTEE survey will form the wide, shallow tier out to intermediate redshifts (z$\sim$0.5) and LADUMA the narrow, deep tier (z$<$1.4). LADUMA will include a deep L-band component plus a deep UHF-band component. The two surveys are therefore highly complementary in that MIGHTEE will observe a larger volume at low redshift and will gather a larger sample of low redshift galaxies, while for the overlapping L-band component, LADUMA's extreme depth will probe lower H{\sc i} masses, and LADUMA's UHF-band component will cover redshifts inaccessible to MIGHTEE.
There is substantial overlap in the science goals of the two surveys,
notably in probing the H{\sc i} mass function (HIMF) and cosmic
neutral gas density over a range of cosmic time and different
environments. At low redshifts MIGHTEE will observe more galaxies at
the high-mass end of the HIMF due to the larger local cosmological
volume probed and LADUMA, due to its $\sim$4.5x higher sensitivity in
the L-band, will detect more low H{\sc i} mass galaxies. By combining
H{\sc i} detections from the two surveys we will be able to measure
both the low- and high-mass ends of the HIMF, significantly reducing the associated statistical errors due to low source counts, out to intermediate redshift \citep[Fig.~\ref{fig:HIMF};][]{Maddox2016}.

\subsection{From the gas to the stars: the onset of star formation
  over cosmic time}

Galaxies follow known scaling relations, such as the so-called
star-formation (SF) main sequence \citep{Johnston2015} relating SF
with stellar mass. Underlying these relations is a complex cycle of
acquisition, storage, consumption, expulsion and re-acquisition of gas
acting to regulate a galaxy's ability to form stars
\citep[e.g.][]{Dave2016}. The vastly different evolution of cosmic SF
density and H{\sc i} density from $z\sim 1$ to the present day implies
a complex, non-linear relation between the two
processes. State-of-the-art simulations which attempt to incorporate a
neutral gas component into galaxies are unable to reproduce the
detailed distribution of H{\sc i} content in galaxies. In particular,
they do not exhibit the mass-dependent relation between H{\sc i} mass
and stellar mass, found to be at least partly due to dark matter halo
angular momentum \citep{Maddox2015}. This mismatch between
observations and simulations indicates missing physics in our galaxy
formation prescriptions that can only be brought to light with large samples of H{\sc i} measurements spanning orders of magnitude in stellar mass.

Furthermore, while the stellar properties of galaxies spanning the
full range of masses over a large range of redshifts have been well
studied, our knowledge of the neutral gas (H{\sc i}) content of these
same galaxies is restricted to the local universe. Surveys such as
ALFALFA \citep{Giovanelli2005} cover large areas of sky and include a
variety of cosmic environments, but do not have the frequency coverage
required to explore a range of redshifts. If we hope to understand the
build-up of stellar mass, we must observe the neutral gas reservoir of
fuel from which the molecular gas forms, eventually turning into
stars, along with the interface between this gas and the galaxies and
environment in which it is located. MIGHTEE will open up this new
parameter space in the study of the neutral gas component of
galaxies. 

MIGHTEE will have the sensitivity and frequency coverage required to detect H{\sc i} in a statistically significant number of M$_{\rm HI}^{\star}$ galaxies at $z=0.2$, and will detect the rare, most H{\sc i}-massive galaxies to $z\sim0.5$, due to the large volume probed. Thus, together with LADUMA, MIGHTEE will revolutionise our understanding of the H{\sc i} content of galaxies over a cosmologically significant redshift range, as a function of stellar mass and environment.

MIGHTEE not only provides us with the ability to trace the origins of
the gas that eventually turns into stars, but with the exquisite
sensitivity to radio-continuum emission, we will also measure the
end-point, namely the star-formation rate (SFR). Radio-continuum
observations closely trace the far-infrared emission of galaxies at
all redshifts, the so-called Far-infrared-radio correlation
\citep[e.g.][]{Jarvis2010,Smith2014}, as such they offer a unique
method of measuring the evolution of the SFR density over cosmic time,
free of dust-obscuration, a key issue in sensitive UV-based surveys
\cite[see e.g.][]{MadauDickinson2014}. Furthermore, deep far-IR and
sub-mm surveys are generally confusion limited at SFR $\sim
100$\,M$_{\odot}$\,yr$^{-1}$ \citep[e.g.][]{Oliver2012,Geach2013}, and
although it provides high angular resolution, ALMA cannot survey large areas efficiently \citep{Dunlop2016}.
At the depth of the MIGHTEE survey, we will be able to detect galaxies
with SFR$\sim 15$\,M$_{\odot}$\,yr$^{-1}$ to $z = 1$ and SFR$\sim
150$\,M$_{\odot}$\,yr$^{-1}$ to $z\sim 3$. By using the excellent
existing optical and near-IR data we are able to trace the evolution
of SFR in galaxies as a function of  mass, colour, environment etc, thus gaining critical insight of which galaxies form the bulk of the stars, in what environments and when.

\subsection{The galaxy merger history from OH megamasers}

The global SFR and the assembly of massive elliptical galaxies are
inextricably connected to how galaxies merge over cosmic time.
Recently merged (luminous and ultra-luminous infrared) galaxies
provide the perfect conditions for detecting Hydroxyl (OH) megamasers,
which are often found within 1\,kpc of heavily dust-obscured AGN
\citep[e.g.][]{2002AJ....124..100D}. OH megamasers are therefore
ideal luminous radio beacons ($L_{\rm OH} \sim 10^{4}\,L_{\odot}$) for
tracing the merger history of the Universe
\citep[e.g.][]{1998A&A...336..815B}. The Arecibo OH Megamaser survey
\cite{2002AJ....124..100D}, which detected 52 masers out to $z =
0.23$, represents the current state-of-the-art in our understanding of
the nearby megamaser population and the luminosity function
\cite{2002ApJ...572..810D}. The MIGHTEE survey will provide a unique opportunity to
carry out a deep blind search for OH megamaser emission (and OH
absorption) between $z = 0$ and 0.85. 
The low-redshift luminosity function of \cite{2002ApJ...572..810D}, would imply a
detection yield of $\sim$10 OH megamasers. However, this number is
highly dependent on the evolution of the galaxy merger rate as a
function of redshift, often parameterised as $(1+z)^{m}$ with $m
\approx 2 - 8$, which could easily lead to an order of magnitude more
in MIGHTEE. With MIGHTEE we will measure the merger rate
traced by OH megamasers directly.

\subsection{Quenching: the role of environment}\label{sec:quenching}

One of the key unknowns in models of galaxy evolution is how SF in
galaxies becomes quenched. Environmentally-related processes such as
ram-pressure stripping and strangulation, where the cold and/or
hot-gaseous haloes are removed due to the density of the medium
through which a galaxy is moving, can explain some of the required
quenching. However, internal processes such as SF and AGN-driven winds
may also play an important, if not critical, role \citep[e.g.][]{Peng2010}. 
Furthermore, it is clear from both semi-analytic models and
hydro-dynamical simulations \citep[e.g.][]{GaborDave2015} and observations \citep[e.g.][]{Hartley2013,Hatfield2016} that these internal feedback processes are also dependent on the mass of the galaxy or halo in which they reside.

Recent work \citep{Hartley2015,Darvish2016,HatfieldJarvis2017} using optical and
near-infrared data has shown that the quenching of low-mass galaxies
in the environments of large massive galaxies, which trace the largest
dark matter haloes, does not appear to be strongly related to their
position in the halo. This suggests that quenching is an internal
process, with the environment playing a lesser role. However, when the
massive tracer galaxy is strongly star forming, then the satellite
galaxies are less quenched. Thus, the process is not simple and
requires an understanding of the evolution of haloes over time, and
also an indication of the time at which the halo itself
formed. Furthermore, it is now clear that the so-called
``Galactic Conformity" \citep[see e.g.][]{Kauffmann2013,Hearin2016}, where neighbouring haloes appear to be related,
continues to high ($z> 2$) redshift \citep{HatfieldJarvis2017}.

The depth of MIGHTEE allows us to directly observe where and when the
cold gas is being stripped from galaxies through the sensitivity to
H{\sc i}, and the areal coverage provides enough cosmic volume to
study such physical processes in the densest and sparsest
environments. With the same set of observations, and 
following the prescription outlined in \cite{HatfieldJarvis2017}, we
will relate the prevalence of quenching to the halo mass, and the proximity to the
centre of the halo, and to the galaxies which may accelerate quenching such as
those exhibiting AGN activity.

\subsection{AGN fueling and feedback}\label{sec:AGN}

It is widely thought that AGN activity may be responsible for
switching off SF in massive galaxies, or at least maintaining a
``quenched" state (Section~\ref{sec:quenching}) once the SF has terminated. 
However, a direct observational link between AGN activity and SF at high redshifts remains elusive. 
Recent studies from both a theoretical \citep{2013ApJ...772..112S} and observational \citep{2014MNRAS.442.1181K} perspective have shown that powerful radio-loud AGN may actually provide a positive form of feedback. On the other hand, there is little evidence for any type of feedback from radio-quiet objects based on studies using \textit{Herschel} \citep[e.g.,][]{2011MNRAS.416...13B}, with recent studies suggesting that the host galaxies of radio-quiet AGN are similar to the general galaxy population \citep[e.g.][]{2013A&A...560A..72R}.

\subsubsection{Tracing the mechanical feedback from AGN jets}

To understand AGN feedback and the interplay between SF and AGN activity, both in the AGN host and the wider environment, a survey is required that spans enough cosmological volume to include the rare powerful AGN at $z\sim 1$, but with a depth that detects SFGs at similar redshifts to the AGN.
% \textit{Herschel} studies are limited due to resolution and confusion noise, and do not provide information about the AGN themselves, whereas optical surveys miss all of the obscured galaxies. Therefore 
Radio is arguably the best line of attack due to the sensitivity to both SF activity and AGN activity. 
% However, current {\em blank-field, wide-area} radio surveys are unable to probe radio emission from SF over the epoch where AGN are having an impact on their environment. 
Given that different forms of AGN feedback are invoked in current semi-analytic and hydrodynamic models of galaxy formation \citep[e.g.][]{2012MNRAS.420L...8H}, it is essential that we understand such processes if we are to understand the evolution of galaxies in general.

Observational evidence \citep[e.g.,][]{2012MNRAS.421.1569B} suggests
that many or most low-power (P $<$ 10$^{25}$\,W\,Hz$^{-1}$) radio
galaxies in the local universe (the numerically dominant population)
correspond to a distinct type of AGN. These sources accrete through a
radiatively inefficient mode (the so-called ``radio mode"), rather
than the radiatively efficient accretion mode typical of radio-quiet
AGN selected at optical or X-ray wavelengths (sometimes called `quasar mode'; see \cite{2014ARA&A..52..589H} for a recent review covering these feedback processes). The role of these two accretion modes appears to be strongly influenced by the environment \citep[e.g.,][]{2008A&A...490..893T} while the level of radio-jet activity  appears to be a strong function of the stellar mass of the host galaxy \citep[e.g.][]{2012A&A...541A..62J,2015MNRAS.450.1538W}. 

Therefore, deeper radio surveys that cover enough area of sky with the best multi-wavelength data are required to probe the evolution of these relationships and the accretion mode dichotomy over cosmic time; this is key information for any attempt to incorporate radiative {\em and} mechanical feedback from radio-loud AGN in models of galaxy, group and cluster formation and evolution.

The depth and breadth of MIGHTEE will enable unique studies of the
entire AGN population from $z\sim0-6$, providing a complete view of
nuclear activity in galaxies and its evolution, unbiased by gas/dust
selection effects. 
If current simulations and measurements are correct, we will detect a factor of $\sim 20$
more low-accretion rate radio sources at $z>1$ than the current
VLA-COSMOS data \citep{Smolcic2017a,Smolcic2017b,Smolcic2017c}, due to the greater depth and increased
area. This dramatic increase means that we will
accurately measure the evolution of such sources to $z\sim 2$, testing
the key ingredient of galaxy evolution simulations. Furthermore, this
will allow the amount of energy deposited into the intergalactic medium
by such objects to be measured over the era when such sources are active, thus providing the key input to the evolution of mechanical feedback from AGN jets.

\subsubsection{\mbox{H\,{\sc i}} as a direct probe of neutral gas accretion and feedback in AGN}

MIGHTEE will provide a homogeneous survey of 21-cm \mbox{H\,{\sc i}} absorption and emission from AGN across the radio luminosity function, enabling a direct investigation into the symbiotic relationship between AGN activity and neutral gas in galaxies. The \mbox{H\,{\sc i}} gas content of AGN has received increasing attention in recent years, both through emission and absorption studies. \mbox{H\,{\sc i}} emission probes the global properties of the neutral gas, such as mass, velocity structure and neutral gas fraction. This offers a direct measurement of the large-scale fuel source for the AGN, and a probe of triggering mechanisms.
Complementary to this, \mbox{H\,{\sc i}} absorption seen against the radio continuum source adds 
vital information on the state and kinematic behaviour of neutral gas within the central regions of radio-loud AGN \citep[e.g.][]{2015A&A...575A..44G}.

Existing samples for both types of study are limited in size, with previous work based on relatively shallow observations of pre-selected (and possibly biased) galaxy samples. MIGHTEE will provide a unique opportunity to revolutionise this field 
through a deep, wide-area dataset, with an unbiased sample of radio-selected AGN and a rich suite of ancillary data not normally available in wider 21-cm surveys.

Using both direct \mbox{H\,{\sc i}} emission detections, and stacking analyses, the \mbox{H\,{\sc i}} gas content of samples of galaxies selected to display AGN activity will be compared against samples of non-AGN matched in mass, environment, and other properties. This will allow the investigation of the origin of the AGN activity and the influence of the AGN on its host galaxy. At the sensitivity of MIGHTEE, it will be possible to detect \mbox{H\,{\sc i}} masses down to around $2 \times 10^{9}\,\mathrm{M}_{\odot}$ out to $z=0.2$ at the $5\sigma$ level, assuming typical line widths of 250 km\,s$^{-1}$. This is a substantial advance in sensitivity over previous studies, and should allow detection of the \mbox{H\,{\sc i}} gas in the AGN hosts, given typical gas fractions of 10\% \citep[e.g.][]{2011MNRAS.416.1739F}. There will be around 1000 AGN of all classes hosted by galaxies more massive than $10^{10}\,\mathrm{M}_{\odot}$ in the MIGHTEE area.

For absorption studies, assuming a typical line width of $100$\,km\,s$^{-1}$, we expect to obtain a sample of approximately 240 radio galaxies brighter than 1\,mJy within the comoving volume bounded by $z_{\rm HI} \approx 0.58$. Approximately 50 of these
sources will be brighter than 10\,mJy and we will be sensitive to cold
gas clouds with peak opacities greater than 1\% and typical column
densities greater than $2 \times 10^{20}\,$cm$^{-2}$. This sample
spans the radio luminosity function between $10^{23} < L_{1.4} / {\rm W\,Hz}^{-1} < 10^{25}$, allowing us to directly observe and test
evolutionary models of neutral gas accretion in low-excitation radio
galaxies \citep[e.g.][]{2016Natur.534..218T}. The brightest radio
galaxies in our absorption sample will also enable a study of jet-mode
feedback in neutral gas \citep[e.g.][]{2013Sci...341.1082M} out to
intermediate cosmological redshifts, and is a natural complement to
the MeerKAT Absorption Line Survey \citep[MALS;][]{MALS}.

\subsection{AGN polarisation as a probe of environment}
Radio-loud AGN are intrinsically highly polarised; the fractional polarisation in high-resolution, high-frequency images can approach the theoretical maximum of $\sim 70$\% \citep{Perley+Carilli96}.

MIGHTEE will provide 
extremely sensitive, broad-band L-band observations, supplemented in
some cases with S-band data as described in Section
\ref{sec:obsstrategy}, and so will give measurements of integrated or moderately resolved polarisation as a function of frequency for many thousands of radio-loud AGN, allowing samples to be subdivided by e.g. luminosity and redshift to measure the evolution of intrinsic and environmental properties. Among other projects, MIGHTEE's polarisation
surveys, coupled with total intensity information and optical
identifications/redshifts for the target sources, will allow us to (i) 
statistically relate observed (de)polarisation to source
 environment, by combining with optical and (where possible) X-ray
data available in the target fields; (ii) investigate the dependence of observed (de)polarisation on
physical size, testing the prediction of models that sources observed at low
resolution become more polarised as they become physically larger,
so that the average Faraday depth to/dispersion in front of the
lobes becomes smaller \citep{Hardcastle+Krause14}; it does indeed
appear to be the cases that sources of larger {\it angular} size
are more polarised in deep surveys \citep{Rudnick+Owen14} but MIGHTEE will give much larger samples and constrain the physical size scale on which this takes place, and hence the physical size of the hot-gas halo; (iii) carry out large-scale statistical studies of the Laing-Garrington effect for a large number of moderately resolved
AGN, probing physical conditions
in the centre of the host environment and giving a prior for
every source on the angle to the line of sight; and (iv) carry out a statistical test of the \cite{Burn66} depolarisation law and hence probe the fine structure and possibly the power spectrum of the magnetic field in the
intra-cluster medium (ICM).

\subsection{Large scale structure \& cosmology}

Covering 20 deg$^2$, MIGHTEE will provide the ideal data set to pave the way for large-scale cosmology experiments 
with the SKA \citep[see ][]{Abdalla2015,Jarvis2015,Santos2015}.

\subsubsection{An accurate measurement of the evolution of bias}
\begin{figure}
\centering
\includegraphics[width=0.32\textwidth]{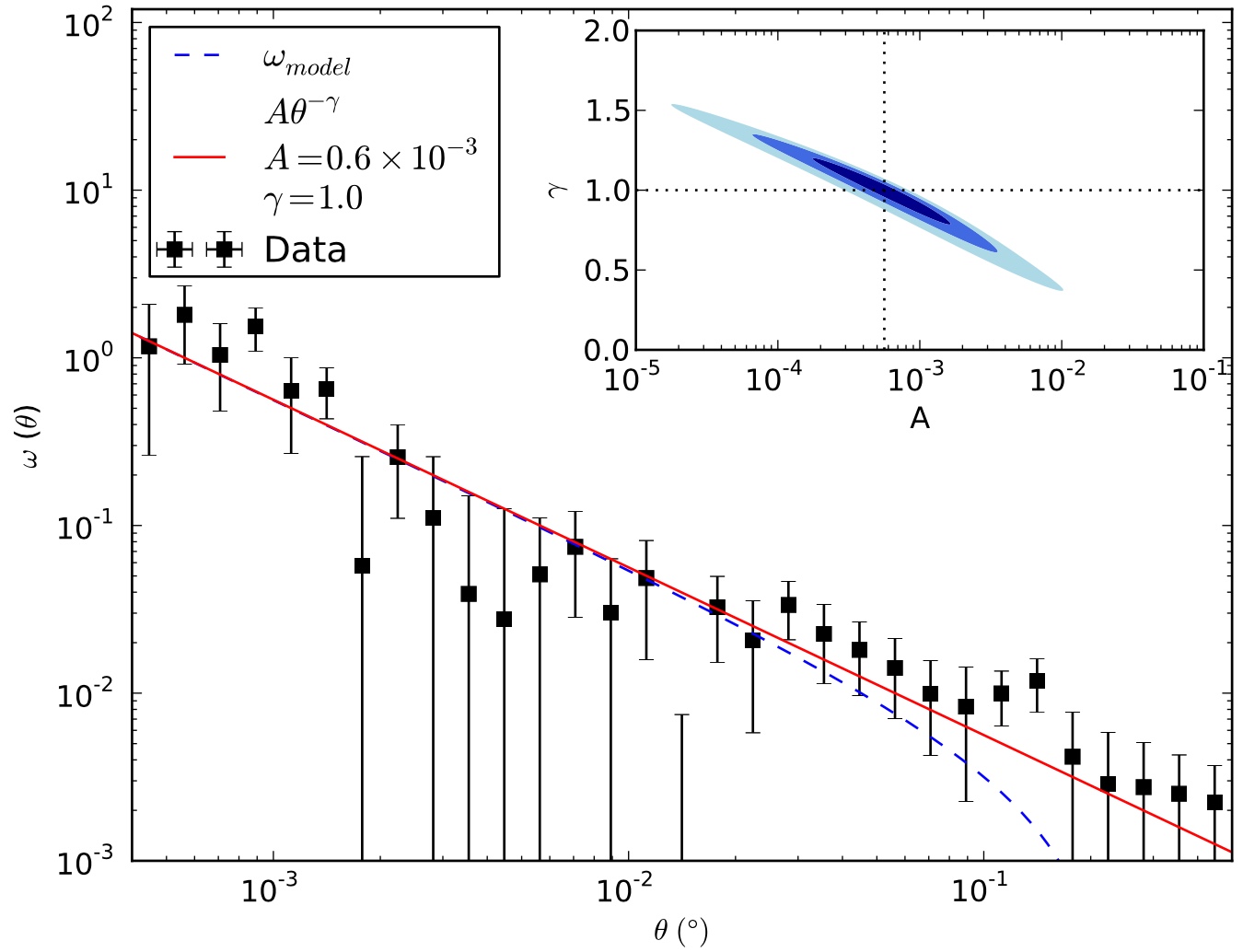}
\includegraphics[width=0.32\textwidth]{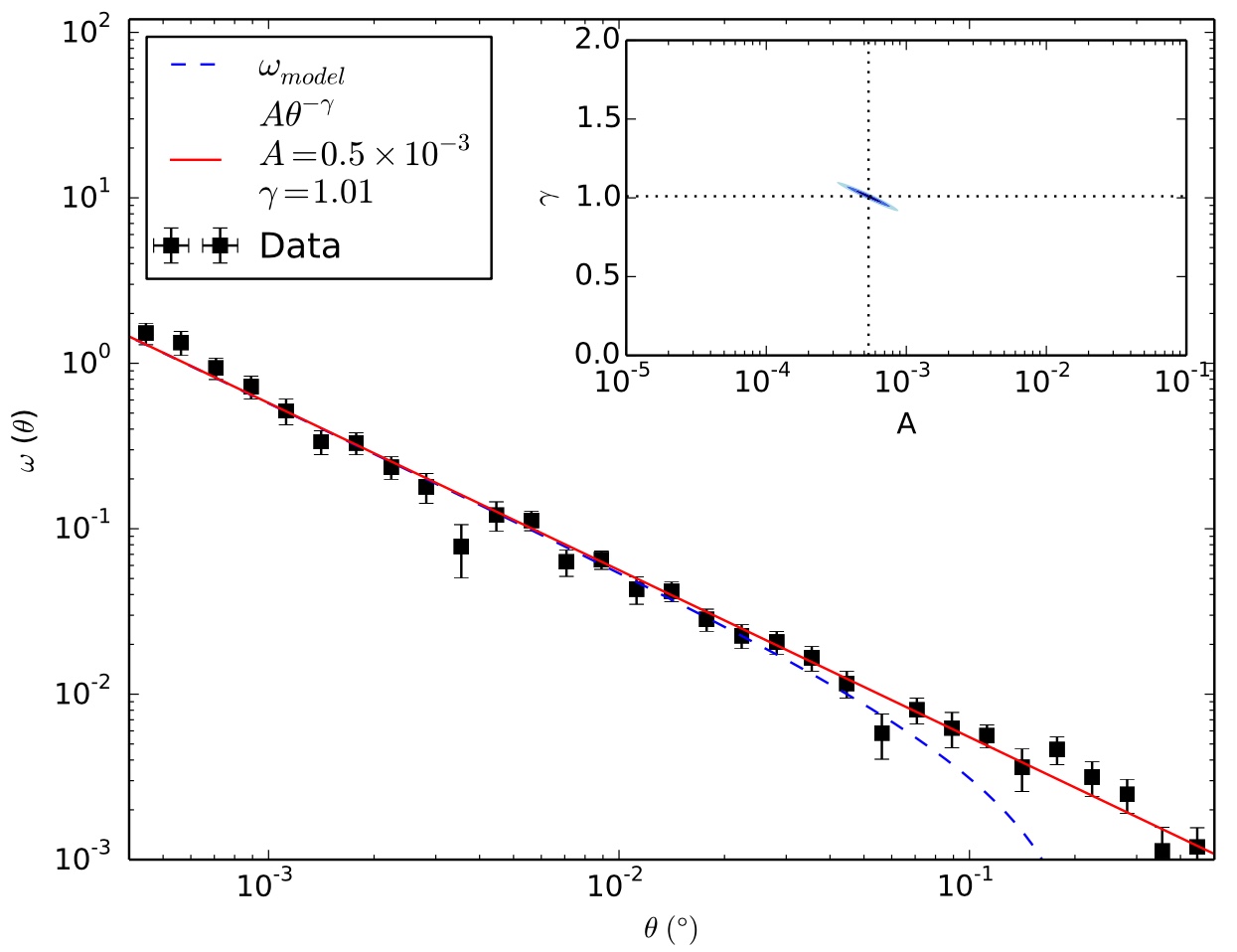}
\includegraphics[width=0.32\textwidth,height=0.25\textwidth]{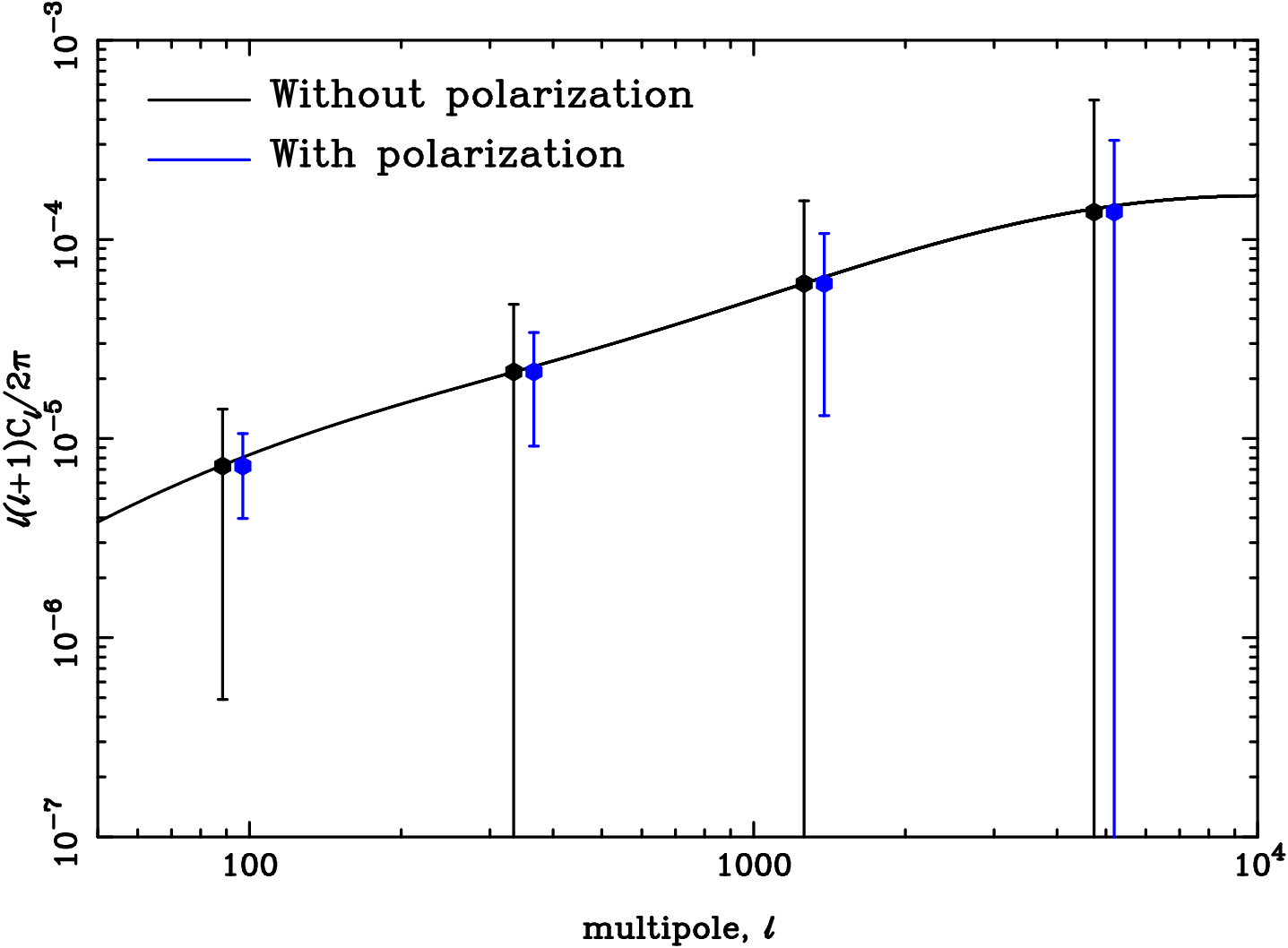}
\caption{\label{fig:TPCF} ({\bf left}) The measured two-point correlation function of radio sources in the recent $\sim 1.5$~deg$^2$ JVLA-COSMOS 3GHz survey (PI: Smol{\v c}i{\'c}), with an equivalent 5$\sigma$ depth of $S_{1.4} \sim 20\mu$Jy (Hale et al. submitted). ({\bf center}) The expected constraints on the two-point correlation function for MIGHTEE with $S_{1.4} = 5\mu$Jy, for a single 7.5\,deg$^{2}$ field.
({\bf right}) Forecasts for a measurement of the cosmic shear weak lensing power spectrum using MIGHTEE polarisation observations combined with galaxy shape measurements from overlapping optical surveys. For comparison we show forecasts for the same sample of galaxies without the polarisation information. We assume we can obtain useful polarisation measurements for $\sim 5\%$ of those galaxies detected in total intensity, and we assume a scatter of $\sim 10$ degrees in the relation between the polarisation orientation and the disk structure of the source galaxies.}
\end{figure}

The depth and breadth of MIGHTEE will allow us to measure the bias,
i.e. how radio sources of different types (e.g. star-forming galaxies,
Fanaroff-Riley Class I and II,
etc), trace the underlying dark-matter distribution \citep[e.g.][]{Lindsay2014a,Allison2015,Mag2017}. Carrying this out over the best multi-wavelength fields allows us to accurately disentangle the different types of sources, whilst obtaining the required number density (apart from very rare FRIIs). Furthermore, cross-correlations with the optical/near-infrared has been shown to be a powerful technique to measure the bias of rarer sources \citep{Lindsay2014b}. Such measurements of the bias can then be used in the wider area surveys where it is much more difficult to measure, due to a lack of multi-wavelength and redshift information. Indeed, these measurements will be needed in order to carry out large cosmological tests with the proposed all-sky SKA-MID survey, as it would be at a similar depth to MIGHTEE. Figure~\ref{fig:TPCF} shows the current state of the art in the measurement of the clustering of radio sources to high-redshift using the most recent JVLA-COSMOS data, and the expected constraints with a single 7.5~deg$^2$ MIGHTEE field. The error ellipse on the amplitude and the slope is improved by around an order of magnitude.

\subsubsection{Weak lensing}

The coherent distortion of distant galaxy shapes due to gravitational light deflection by large scale structure (``weak lensing" or ``cosmic shear") is recognised as one of the most powerful probes of dark matter and dark energy, and is a major cosmology science goal for the SKA \citep{2015aska.confE..23B,Harrison2016,Bonaldi2016}. 

MIGHTEE does not have the resolution to measure shapes of galaxies and thus cannot directly measure the shear for weak lensing measurements. However, by conducting the survey over the fields with the best optical data (which have been and continue to be used for weak lensing measurements), we are able to use the integrated polarisation vector from MIGHTEE data, to inform on the intrinsic alignment of the galaxies before they are lensed. This removes one of the most significant systematics in weak lensing surveys, and if proven would allow MIGHTEE to be the path finder for weak lensing with polarisation for the SKA.

Our forecasts suggest that MIGHTEE polarisation data combined with shape information from best overlapping optical data, will provide a $\sim 3.5\sigma$ detection of the cosmic shear effect (Fig.~\ref{fig:TPCF}; right). This would represent the first detection of weak lensing using the polarisation technique.

\subsubsection{Cosmic magnetism and large-scale structure: The magnetic cosmic web}
MIGHTEE will provide a supremely-dense Rotation Measure (RM) grid 
- an effect caused by the interaction of a
magneto-ionic medium, with the linearly polarised synchrotron emission
from background or embedded cluster galaxies - 
and permit measurement of the properties of magnetic fields embedded in the large-scale structure of the universe. 
While clusters contain hot plasma of $T>10^7$\,K,  filaments are expected to be filled with plasma of $10^7$ K $>$ T $>10^5$ K, referred to as the Warm Hot Intergalactic Medium (WHIM). The plasmas may be magnetised; diverse processes for seed magnetic fields have been suggested, and the seed 
 fields can be further amplified through compression and turbulent dynamos as well as leakage of galactic media during the hierarchical structure formation in the universe \citep{2012SSRv..166....1R}.
 
 Simulations predict that the inter-galactic magnetic field (IGMF) in filaments would induce excess Faraday Rotation Measures with a flat second-order structure 
 function of $\sim$100 rad$^2$\,m$^{-4}$ for angular separation of $r \gtrsim 0.1^{\circ}$.
The power from this contribution to the RM structure function will be distinguishable from the foreground of the Milky Way galaxy on angular scales smaller than a few degrees.
 An RM data set with a sky density of several 100 to 1000 polarised sources per square degree, and with RM precision of $\sim$1 rad~m$^{-2}$ is required to accurately reconstruct the structure function of RM
 variance due to the IGMF in the cosmic web \citep{2014ApJ...790..123A}.
MIGHTEE will probe this structure function with the required number density and RM precision, and over angular scales of a few arcminutes to several degrees, offering our best opportunity before the SKA to use this technique to detect the magnetic cosmic web.

\subsubsection{Resolving Massive Galaxies out to Cosmological Distances: Dark Matter in Galaxies}

MIGHTEE's combination of high sensitivity and angular resolution for H{\sc i} produces a  column density sensitivity ($\sim$1 M$_{\odot}$/pc$^2$ over $\delta_V$ = 20 km/s at S/N=3) that is sufficient to resolve the H{\sc i} distributions of H{\sc i}-rich galaxies out to cosmological distances. 
MIGHTEE will therefore be the first survey to probe the distribution of dark and luminous matter in a statistical sample of massive galaxies beyond the local volume.
The tight correlation between H{\sc i} mass and H{\sc i} diameter for local galaxies \citep{Wang2016}, coupled with the expected number of H{\sc i} detections using the local HIMF \citep{Martin2010},
implies that MIGHTEE will resolve $>150$ galaxies with M$_{\rm HI}$ $>$ 10$^{10}$ M$_{\odot}$ at redshifts $0.07 < z < 0.12$. 
This is sufficient to estimate the H{\sc i} distribution and rotation curve shape by modelling the 3D H{\sc i} cubes and optical images \citep{Kamphuis2015}.  
MIGHTEE is the only planned H{\sc i} survey capable of resolving
a significant sample of massive galaxies out to cosmological distances before the advent of SKA1, 
and highly complementary to samples at lower redshifts that will be
produced by wide-field H{\sc i} surveys with the Australian Square
Kilometre Array Pathfinder \citep[ASKAP;][]{Johnston_ASKAP} and \citep[Apertif;][]{APERTIF}. 

\subsubsection{Low-mass, nearby galaxies and the satellite problem}

Predictions based on $\Lambda$CDM excel at matching observations of
large scale structure. However, on small scales, the effects of
baryonic physics are important and it can be difficult to match
simulations to observations. Observations of low-mass galaxies offer
constraints on the implementation of baryonic physics, and are especially valuable as they include
spatially resolved kinematic information. Systems with H{\sc i} masses below $\sim$10$^8$ M$_{\odot}$ are especially important for testing our understanding of baryonic physics and $\Lambda$CDM. The H{\sc i} kinematics of galaxies in this mass range place them in dark matter halos incompatible with expectations from abundance matching results \citep{Papastergis2015}. 
The lowest mass systems (below $\sim 2 \times 10^7$ M$_{\odot}$) are extremely rare; in the full ALFALFA survey there are only $\sim$70. 
Finding and studying these lowest mass galaxies is critical for understanding at what mass range and due to which processes dark matter halos stop hosting observable galaxies. With MIGHTEE we 
% calculate the that the 5-sigma detection limits for a 30 km/s line width source (typical for galaxies with masses below a few x 10$^7$ M$_{sun}$) is 0.0112 Jy km/s for a perfectly matched channel size. Based on the ALFALFA HIMF,
expect to detect $\sim$270 sources with M$_{\rm HI} < 10^8$ M$_{\odot}$, with $\sim$15 of those below 10$^7$ M$_{\odot}$. Using the HI mass-diameter relation \citep{Wang2016}, we estimate that $\sim$7 of the sources below 10$^8$ M$_{\odot}$ will be resolved with at least three MeerKAT beams, and one to two sources may be resolved with five beams.% This will increase our census of low mass galaxies and the number of systems where the hosting dark matter halo may be constrained. 

\subsection{Cluster radio halos and relics}
Clusters of galaxies show radio emission on a very broad range of
scales, from discrete sources associated with AGN to diffuse emission on Mpc scales.
The latter points to the existence of a non-thermal component (cosmic rays and magnetic fields) of the ICM. 
Diffuse radio sources at the cluster center are generally known as ``radio halos'', while elongated polarised
Mpc-scale radio structures at the cluster periphery are known as ``radio relics''. 
 Current models
indicate that Mpc-scale diffuse radio emission in the form of radio
halos traces the turbulent re-acceleration of particles due to merger
events, whereas relics are the result of electron acceleration and
magnetic field compression resulting from ICM shocks \citep[e.g.][]{2014IJMPD..2330007B}.

The angular resolution and excellent brightness sensitivity
provided by MIGHTEE will probe radio emission on all scales relevant
for cluster radio emission to high redshift.  MIGHTEE's multi-wavelength coverage will be crucial in shedding
light on the origin of the non-thermal cluster components, as detailed
multi-wavelength analyses of clusters are essential to properly
characterise the physical differences between ``radio-loud" and
``radio-quiet" systems. This kind of study is crucial to
understand, for instance, why not all merging clusters host diffuse
radio sources, or vice-versa \citep[e.g.][]{Bonafede2014}, and the role played by gravitational
and non-gravitational effects on the evolution of the ICM.

\subsection{Galaxy clusters and their magnetic fields}

Galaxy clusters are the largest known magnetised structures in the universe and therefore are unique laboratories to investigate the origin of large-scale magnetic (B) fields. 
In spite of the wealth of evidence for the existence of ICM magnetic fields, measurements of both the field strengths and morphologies of clusters are still scattered within the literature. 
Information on the cluster B-fields can be derived from detailed
images of extended radio emission in clusters \cite[radio halos,
mini-halos and relics;][]{Tribble1991} and through Faraday Rotation
measure synthesis. With MIGHTEE we will be able to measure the magnetic field properties of hundreds of galaxy clusters in the MIGHTEE cosmic volume.
The Faraday rotation provides a clean measure of the magnetic field strength along the line of sight for uniform fields, and the magnetic field strength directly for turbulent fields \citep{1982ApJ...252...81L}.

Constraining the behaviour of magnetic fields in galaxy clusters is
not only important for studies of the intra-cluster medium itself, but
can also be used to constrain the origin of cosmic magnetic fields
more widely by examining the evolution of cluster fields as a function
of redshift. Early-type seed fields are expected to produce a
characteristic evolution in cluster magnetic field properties when
combined with expectations of turbulent amplification. The high
density and precision of the MIGHTEE RM grid will offer the best
opportunity before SKA to constrain the redshift evolution of cluster
magnetic fields, and provide the first test of the early-type model for the origin of cosmic magnetic fields.

\subsection{The emergence and evolution of magnetic fields in galaxies}
Through its ability to detect polarisation of sources at high redshift, 
MIGHTEE will be a cornerstone for investigating the evolution of magnetic fields in galaxies. 
For disk galaxies, this range is $z \lesssim$ 2.5, and for AGN and starbursts it reaches out to $z \lesssim$ 7, into the epoch of reionisation. 
Over this period, galaxies formed and evolved, converting most of their 
gas into stars. 
The evolution of plasma and magnetic fields in 
galaxies is expected to be closely related to the evolution of the cosmic SFR, and tied to the intergalactic medium through
accretion, galactic winds, tidal and ram stripping, and AGN activity.
While $\mu$G strength 
magnetic fields on kpc scales may have formed in galaxy disks by $z \sim 3$,  ordering on the scale of a galaxy may have taken until 
$z \sim 0.5$, depending on galaxy mass \citep{2009A&A...494...21A,Mao2017}.
Competing with this, galaxy interactions and continuous feedback by supernovae and stellar winds can 
enhance the turbulent component of the magnetic field, and drive outflows that transport plasma and magnetic field 
from the disk into the halo. Since these processes scale with the global SFR, significant evolution is expected 
between $z \sim 2$ and the present. 
Also, the density of Faraday rotating plasma will gradually decrease over time as galaxies 
transform a significant fraction of their gaseous mass into stars, implying a gradual evolution in Faraday depth.
 
\begin{figure}[h]	
\centering 
\includegraphics[width = 0.95\textwidth]{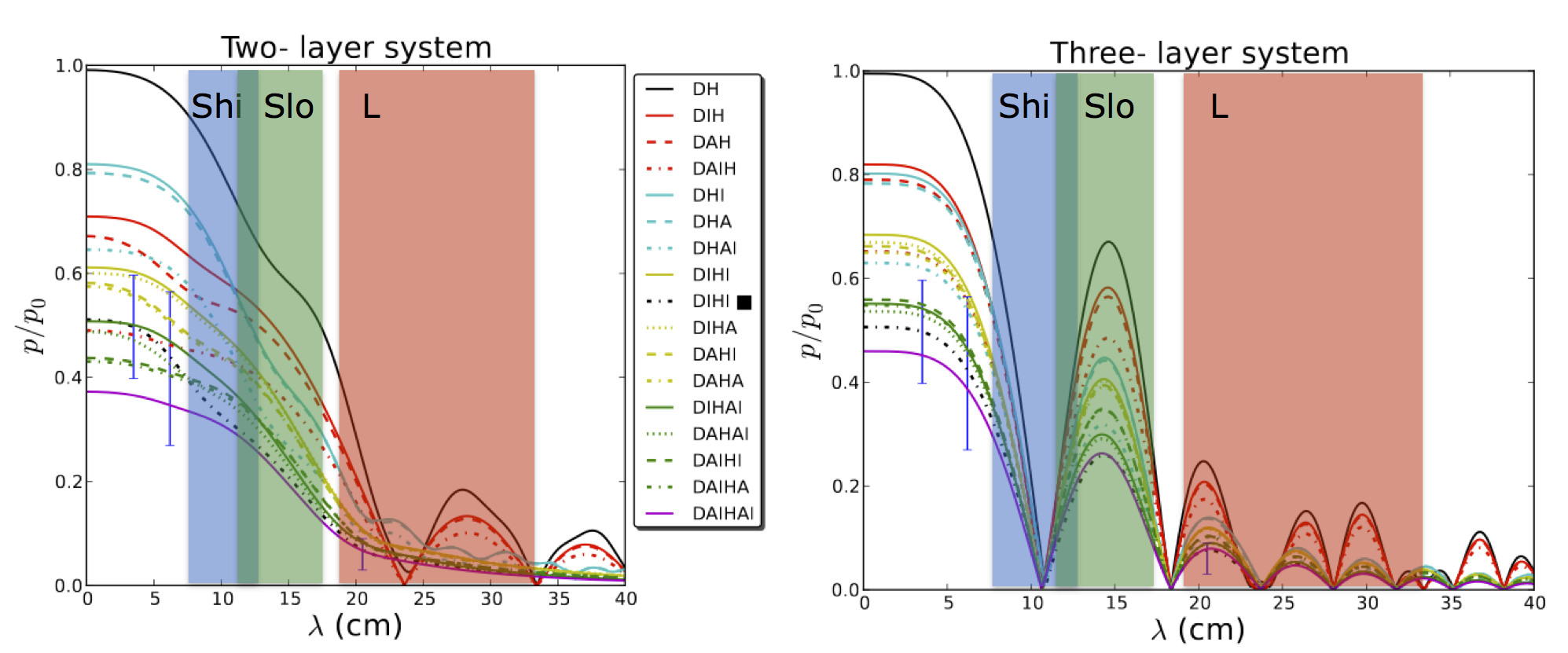} 
\caption{Models of the polarised intensity of
integrated emission from galaxy disks as a function
of frequency. The frequency coverages of the MeerKAT 
L-band and S-band are shown.  
Together these bands span the frequency range over which depolarisation becomes important, and both are critical to measure the depolarisation signature. 
The change in polarised emission is due to internal 
Faraday effects from the magnetic field and thermal plasma in the galaxy. 
Broad-band integrated spectro-polarisation data thus probes the evolution of both magnetic and thermal gas (ionisation) properties. Depolarization models are
from \cite{2014A&A...567A..82S}. }
\label{fig:depol} 
\end{figure}

Based on models by \cite{2009ApJ...693.1392S} for spiral galaxies at
$z =0$, applied to normal SF  galaxies in the SKADS Simulated Skies
simulation \citep{Wilman2008,Wilman2010},
at the sensitivity of the MIGHTEE deep commensal 
polarisation image of the LADUMA field we expect $\sim$5000 galaxies per square degree above 10$\sigma$ detected to $z\sim 3$ .  
For the wider area survey we will detect
several hundred galaxies per square degree in polarisation out to
$z \sim 1$.
The analysis of \cite{2009ApJ...693.1392S} has shown that
the integrated polarisation properties of these
distant galaxies can be used to detect and characterise both the large-scale and turbulent magnetic fields 
in galaxy disks. 
As shown in Figure~\ref{fig:depol}, the combined L- and S-band data set is critical to this study as the 
structure of the depolarisation of the signal as a function of frequency over this broad band is required
to reveal the magnetic properties. 
The MIGHTEE data set will provide probes of the emergence and evolution of magnetic fields 
in $\sim$10,000 galaxies out to $z > 3$.

\subsection{Digging into the noise with MIGHTEE}
Classical analyses of radio survey data relies on the direct detection of the H{\sc i},
continuum or polarised emission from galaxies. 
However, there is significant additional information in the images for the
vast number of objects that produce radio signals too faint to individually reach the direct detection flux density threshold.
We will be able to take
advantage of this information and investigate the statistical radio properties of classes of objects
to flux densities substantially below the noise floor of the images.
Following techniques developed in
\cite{Mitchell-Wynne2014} and \cite{Zwart2015} we plan to derive deep
H{\sc i} mass functions, and total intensity and polarised
luminosity functions through a Bayesian stacking method. Information
from the rich multi-wavelength data set for the MIGHTEE fields can be used as prior information in this
process.  For example, $\sim 200,000$ galaxies per square
degree, based on the VIDEO survey data \citep{Jarvis2013} to $K_s \sim 23.7$, would result
in a total of $4\times 10^6$ galaxies to stack on over all redshifts
and stellar masses.  
While ultimately limited by confusion in total intensity, MIGHTEE will provide an unique opportunity: due to its high sensitivity, a
large number of the MIGHTEE galaxies detected in continuum will be
SFGs containing H{\sc i}, and for which we can also measure their magnetic fields through the stacked polarised emission that is unaffected by confusion.

\section{MIGHTEE observations and data}
\subsection{Observing strategy}\label{sec:obsstrategy}

\begin{figure}	
\centering 
\includegraphics[width = 0.9\textwidth]{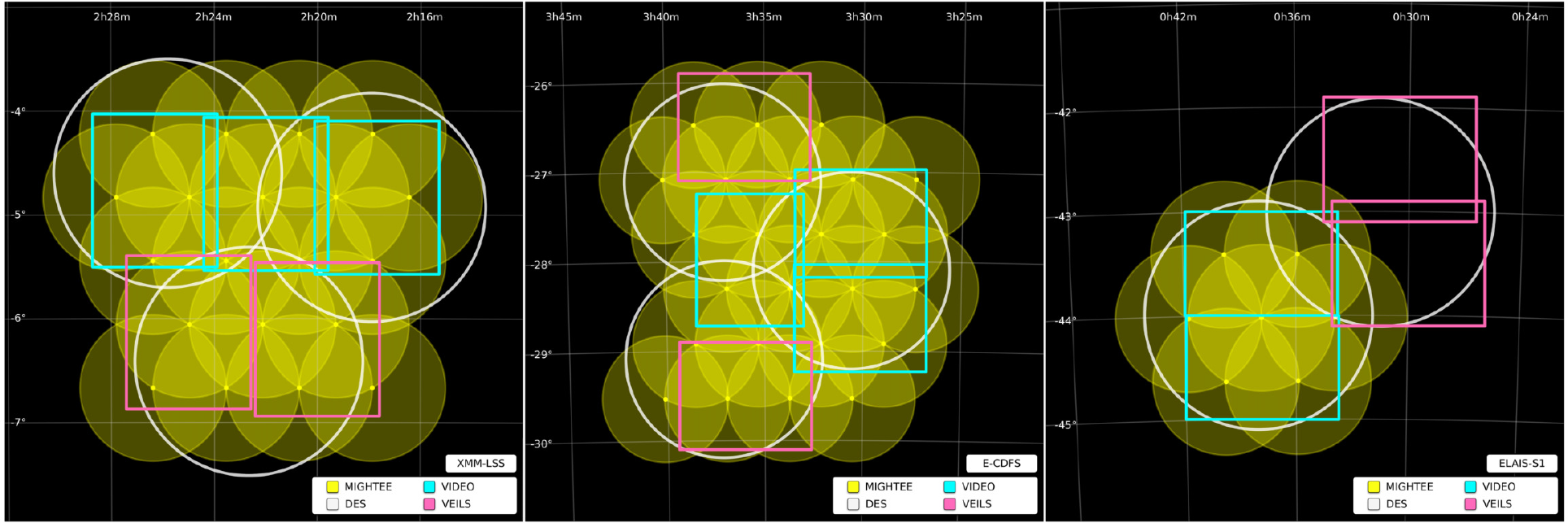} 
\caption{Current plausible pointing strategies for (left-to-right) XMM-LSS (20 pointings, 6.7 deg$^{2}$), E-CDFS (24 pointings, 8.3 deg$^{2}$) and ELAIS-S1 (7 pointings, 1.6 deg$^{2}$). Not shown here is the fourth COSMOS field, which will consist of a single deep pointing. In practice the grid of E-CDFS pointings will be snapped to the LADUMA pointing centre, requiring only 23 additional pointings from MIGHTEE.}
\label{fig:ptgs} 
\end{figure}

The key science outlined above, coupled with  the current and
anticipated availability of multi-wavelength data, lead to the mosaic pointing setups shown in Figure \ref{fig:ptgs}. 
The number of pointings and the areas (calculated without including the area beyond the half
power point of the perimeter pointings) are provided in the
caption. Not shown is the COSMOS field which will be observed with a
single (or tightly dithered) pointing. Using the current measurements of the MeerKAT
system temperature,  with 16\,h per pointing a depth of
2\,$\mu$Jy~beam$^{-1}$ (thermal + confusion noise) will be achieved in
the full-band mosaics, and a typical depth of 90\,$\mu$Jy~beam$^{-1}$
will be reached in the 26\,kHz channels for the spectral line component
of the survey. We note that one of the E-CDFS pointings will be that of the MIGHTEE-DEEP tier commensal with LADUMA.

The S-band component will cover 4 deg$^{2}$ in E-CDFS and
1.5~deg$^{2}$ in COSMOS  (we do not plan to survey XMM-LSS in S-band due to satellite RFI observed in test JVLA observations). 
The S-band survey over a limited area of MIGHTEE will enable us to obtain a much larger bandwidth for RM synthesis, 
while also allowing the possibility of multi-frequency synthesis in
combination with the L-band,  which will help to deconfuse the 
L-band survey.
We plan to reach a depth of 1\,$\mu$Jy~beam$^{-1}$ in the S-band
mosaics (matched to the L-band depth for a typical
$\alpha =-0.7$ source), which requires 12.7\,h per pointing (ignoring the
effects of confusion, but including the overlapping pointings, and
assuming a 20\% sensitivity loss due to the weighting required for
reliable deconvolution).  Calibration overheads of $\sim$20\% means that the total time required is $\sim$1920~h.

\subsection{MIGHTEE scientific data products}
We plan to provide multiple intermediate MIGHTEE data releases of the reduced data following qualification of the data 
products for scientific use.  Data will be released along with ancillary data where appropriate. 
Visibility data will be processed into images and catalogues using the data-centric processing facility at the 
South African Inter-University Institute for Data Intensive Astronomy (IDIA).
The data processing pipeline is under development by a coordinated development team using the IDIA cloud-based 
development framework. 
We envisage that the MIGHTEE team will provide a data release 18 months after observations of a field are
completed in either L- or S-band. 
The release will include data cubes for spectral line work,  band-averaged multi-frequency synthesis total intensity images, full-Stokes intensity and RM synthesis cubes. Continuum products will have multiple resolutions optimised for 
different science goals, e.g. high resolution for galaxy studies, and lower resolution for cluster halo/relic work.
We plan to release these data through the IDIA systems and are also investigating plans for a mirrored archive in Europe.
Final data products will be released to the MeerKAT legacy data archive.

\subsection{VLBI enhancement of MIGHTEE science}

MeerKAT's data alone will be limited to an angular resolution of a few
arcseconds. The $\mu$Jy radio sources will therefore be unresolved in the MIGHTEE images.
This faint population is known to consist of emission from SFG and RQ AGN.
Although multi-frequency analysis can aid in distinguishing the two populations,
higher resolution radio imaging would allow unambiguous distinction through morphology
as well as precise measures of the relative contribution of SF and AGN activity in individual objects.
Very Long Baseline observations would thus offer a tremendous enhancement to the MIGHTEE project.
VLBI can also play an important role in H{\sc i} absorption
components of MIGHTEE, contributing unique morphological insights to
the interpretation of gas inflows and outflows in particular \citep[e.g.][]{2013Sci...341.1082M}. VLBI polarimetry would not only
provide detailed information on sub-kpc magnetic fields and jet
physics, but also enable comparison with (and separation from) the
larger scale polarisation properties to be probed by MeerKAT
\citep[e.g.][]{2015aska.confE..93A}. Over the course of the MIGHTEE
survey the project team  will explore the possibility of including MeerKAT along with elements of the emerging
Africa VLBI Network and European antennas in complementary wide-field VLBI experiments to provide
high-resolution radio images of MIGHTEE sources.

\bibliography{mightee}

%\end{thebibliography}
\appendix
\section*{Author affiliations}
\noindent\tiny{$^{1}$Astrophysics, University of Oxford, Denys Wilkinson Building, Keble Road, Oxford OX1 3RH, UK\\
$^{2}$Department of Physics and Astronomy, University of the Western
Cape, Robert Sobukwe Road, Bellville 7535, South Africa\\
$^{3}$Department of Astronomy, University of Cape Town, Rondebosch 7701, South Africa\\
$^{4}$Inter-University Institute for Data Intensive Astronomy, University of Cape Town, South Africa\\
$^{5}$Instituto de Astrof\'{i}sica de Andaluc\'{i}a (CSIC), Apartado 3004, E-18080 Granada, Spain\\
$^{6}$CSIRO Astronomy and Space Science, PO Box 76, Epping NSW 1710,
Australia\\
$^{7}$Department of Physics \& Electronics, Rhodes University, PO Box 94,
Grahamstown, 6140, South Africa\\
$^{8}$Inter-University Centre for Astronomy and Astrophysics, India\\
$^{9}$ASTRON, the Netherlands Institute for Radio Astronomy, Postbus 2, 7990AA, Dwingeloo, The Netherlands\\
$^{10}$Square Kilometre Array South Africa, Pinelands 7405, Cape
Town, South Africa\\
$^{11}$Jodrell Bank Centre for Astrophysics, School of Physics and Astronomy,
The University of Manchester, Oxford Road, Manchester M13 9PL, UK\\
$^{12}$INAF - Istituto di Radioastronomia, via Gobetti 101, 40129 Bologna, Italy\\
$^{13}$Institute of Cosmology \& Gravitation, University of Portsmouth, Dennis Sciama Building, Portsmouth PO1 3FX, UK\\
$^{14}$Department of Physics and Astronomy, Rutgers, The State
University of New Jersey, 136 Frelinghuysen Road, Piscataway, NJ
08854-8019, USA\\
$^{15}$ African Institute for Mathematical Sciences, 6-8 Melrose Road, 
Muizenberg 7945, South Africa\\
$^{16}$Department of Maths and Applied Maths, University of Cape Town,
Cape Town, South Africa\\
$^{17}$South African Astronomical Observatory, Observatory, Cape Town, 7925, South Africa\\
$^{18}$SUPA, Institute for Astronomy, Royal Observatory, Edinburgh EH9
3HJ, UK\\
$^{19}$Hamburger Sternwarte, Universit\"{a}t Hamburg, Gojenbergsweg 112, D-21029 Hamburg, Germany\\
$^{20}$School of Physics, University of the Witwatersrand, Private Bag
3, 2050 Johannesburg, South Africa\\
$^{21}$CHPC, CSIR, 15 Lower Hope Rd, Rosebank, 7700, South Africa\\
$^{22}$Laboratoire Lagrange, Universit\'{e} C\^{o}te d'Azur, Observatoire de la C\^{o}te d'Azur, CNRS, Blvd de l'Observatoire, CS 34229, F-06304 Nice cedex 4, France\\
$^{23}$Centre for Astrophysics Research, School of Physics, Astronomy and
Mathematics, University of Hertfordshire, College Lane, Hatfield AL10
9AB, UK\\
$^{24}$Max-Planck-Institut f\"ur Radioastronomie
Auf dem H\"ugel 69, 53121 Bonn, Germany\\
$^{25}$European Southern Observatory, Karl-Schwarzschild-Str. 2, 85748, Garching, Germany\\
$^{26}$Astrophysics \& Cosmology Research Unit, School of Mathematics, Statistics \& Computer Science, University of KwaZulu-Natal, Durban 4041, South Africa\\
$^{27}$Western Sydney University, Locked Bag 1797, Penrith South, NSW
1797, Australia\\
$^{28}$Astronomy Centre, Department of Physics and Astronomy, University of Sussex, Falmer, Brighton, BN1 9QH, UK\\
$^{29}$ Centre for Space Research, North-West University, Potchefstroom
2520, South Africa\\
$^{30}$Leiden Observatory, Leiden University, P.O. Box 9513, NL-2300 RA Leiden, The Netherlands\\
$^{31}$International Centre for Radio Astronomy Research, Curtin 
University, Perth, Australia\\
$^{32}$Gemini Observatory, Northern Operations Center,
670 North A‘ohoku Place, Hilo, HI 96720-2700, USA\\
$^{33}$Department of Physics, Royal Military College of Canada, PO Box 
17000, Station Forces, Kingston, ON K7K 7B4, Canada\\
$^{34}$Department of Physics and Astronomy, University of Calgary,
Canada 0000-0003-2623-2064\\
$^{35}$GEPI, Observatoire de Paris, CNRS, Universite Paris Diderot, 5 place Jules Janssen, F-92190 Meudon, France\\
}

\end{document}